\documentclass[]{article}  
\RequirePackage{graphicx}
\usepackage{pstricks,pst-coil}
\begin{document}

\title{ \textbf{Baryon production in the quantized fragmentation of helical QCD string}}


\author{\v{S}\'{a}rka Todorova-Nov\'{a}\footnote{e-mail:sarka.todorova@cern.ch}    }

\maketitle

\begin{abstract}
     Baryon production is studied within the framework of quantized
     fragmentation of QCD string.  Baryons appear in the model
     in a fairly intuitive way,  with help of causally connected
     string breakups.  A simple helical approximation of QCD
     flux tube with parameters constrained by mass spectrum
     of light mesons is sufficient to reproduce masses of light baryons. 
     
\end{abstract}

\section{Introduction}
\label{intro}
 The idea to model the confinement with help of helical string has
 been proposed by Lund theorists in~\cite{helix}.  Causal approach to the ordering of string
 breakup vertices  reveals that mass spectrum of light hadrons
  corresponds to breaking of string in pieces with quantized
 (helix) phase difference~\cite{helix2}. 

 For better understanding of the difference between the Lund string
 model~\cite{lund}, widely used by the particle physics community, and its 
 quantized helical spin-off,  a brief overview of definitions and assumptions
 is provided.

  In close analogy with Lund string fragmentation, the momentum
  acquired by massless quarks moving along the helical string amounts to the integral of the string tension tangential to
  the string curvature. The non-trivial part of the integral between
  vertices A,B is the
  transverse momentum \footnote{terms ``transverse'' and
    ``longitudinal''  describe orientation with respect to string axis}
\begin{eqnarray}   \label{eq:calc}
   \vec{p}_T & = & \kappa R \int_A^B\exp{i(\Phi\pm\pi/2)} d\Phi     \nonumber  \\
                  & = & 2 \kappa R \sin{\frac{\Phi_B-\Phi_A}{2}}  \exp{i(\frac{\Phi_A+\Phi_B\pm\pi}{2}) },
\end{eqnarray} 
where $\kappa \sim$1 GeV/fm stands for
 string tension,  $R$ is the radius of helix and $\Phi$ designs the
 helix phase. The $\pm$ sign distinguishes movement of quarks and
 antiquarks.  The transverse and longitudinal
 momentum components decouple, and quark acquires an effective mass
 entirely derived from transversal properties of the string
\begin{equation}    \label{eq:mass}
   m_{AB} = \kappa R \sqrt{(\Phi_B-\Phi_A)^2 - ( 2\sin{\frac{\Phi_B-\Phi_A}{2}} )^2     }.
\end{equation}
   The 3-dimensional string plays  essential role in the model by allowing to
   introduce causality in the description of the fragmentation
   process -  the use of 1-dimensional string to model the confinement
   in the standard Lund fragmentation model implies that 
   breakup-vertices forming a hadron are - by construction! - causally
   disconnected. 
   
\begin{figure}[t]
\begin{pspicture}(0,-2.5)(10,2.5)
\psline[linewidth=0.1,linecolor=green,linestyle=dotted]{->}(0,2.3)(4.5,2.3)
\psline[linewidth=0.1,linecolor=green]{->}(0,2.3)(1.,2.3)
\psline[linewidth=0.1,linecolor=green]{-}(1.8,2.3)(2.65,2.3)
\psline[linewidth=0.1,linecolor=green]{->}(3.55,2.3)(4.5,2.3)
\pscoil[linewidth=0.04,coilwidth=0.2,coilheight=1,coilarm=0.05]{}(0.5,2.3)(0.5,1.5)
\psline[linewidth=0.04]{->}(0,1.)(0.25,1.25) \psline[linewidth=0.04]{-}(0.,1.)(0.5,1.5)
\psline[linewidth=0.04]{->}(0.5,1.5)(1.,1.)
\psline[linewidth=0.04,linestyle=dotted]{-}(1.,1.)(1.5,0.5)
\psline[linewidth=0.04]{->}(1.5,0.5)(2.1,-0.1)
\psline[linewidth=0.04]{-}(2.4,-0.1)(3.,0.5) \psline[linewidth=0.04]{->}(2.4,-0.1)(2.7,0.2)
\pscoil[linewidth=0.04,coilwidth=0.2,coilheight=1,coilarm=0.05]{}(4.,2.3)(4.,1.5)
\psline[linewidth=0.04]{->}(3.5,1.)(3.75,1.25) \psline[linewidth=0.04]{-}(3.5,1.)(4.,1.5)
\psline[linewidth=0.04,linestyle=dotted]{-}(3.5,1.)(3.,0.5)
\psline[linewidth=0.04]{->}(4.,1.5)(4.5,1.)
\pscoil[linewidth=0.04,coilwidth=0.2,coilheight=1,coilarm=0.05]{}(1.75,2.3)(1.75,0.3)
\pscoil[linewidth=0.04,coilwidth=0.2,coilheight=1,coilarm=0.05]{}(2.75,2.3)(2.75,0.3)
\pscircle[linewidth=0.04,fillcolor=gray,fillstyle=solid](2.25,-0.2){0.2}
\psline[linewidth=0.1,linecolor=green,linestyle=dotted]{->}(5.5,2.3)(10,2.3)
\psline[linewidth=0.1,linecolor=green]{->}(5.5,2.3)(6.5,2.3)
\psline[linewidth=0.1,linecolor=green]{-}(7.,2.3)(8.3,2.3)
\psline[linewidth=0.1,linecolor=green]{->}(8.8,2.3)(10.,2.3)
\pscoil[linewidth=0.04,coilwidth=0.2,coilheight=1,coilarm=0.05]{}(6.,2.3)(6.,1.5)
\psline[linewidth=0.04]{->}(5.5,1.)(5.75,1.25) \psline[linewidth=0.04]{-}(5.5,1.)(6.,1.5)
\psline[linewidth=0.04]{->}(6.,1.5)(6.5,1.)
\psline[linewidth=0.04,linestyle=dotted]{-}(6.5,1.)(6.7,0.8)
\psline[linewidth=0.04]{-}(6.7,0.8)(6.9,0.6)
\pscoil[linewidth=0.04,coilwidth=0.2,coilheight=1,coilarm=0.05]{}(6.9,0.6)(7.7,0.6)
\psline[linewidth=0.04]{->}(6.9,0.6)(6.9,-0.4)
\psline[linewidth=0.04]{->}(7.8,0.)(8.3,-0.5)
\psline[linewidth=0.04]{-}(7.3,-0.5)(7.8,0.)
\psline[linewidth=0.04]{->}(7.3,-0.5)(7.55,-0.25)
\pscoil[linewidth=0.04,coilwidth=0.2,coilheight=1,coilarm=0.05]{}(9.,2.3)(9.,-1.2)
\psline[linewidth=0.04,linestyle=dotted]{-}(8.3,-0.5)(8.8,-1.)
\psline[linewidth=0.04]{->}(8.8,-1.)(9.5,-1.7)
\pscoil[linewidth=0.04,coilwidth=0.2,coilheight=1,coilarm=0.05]{}(7.8,2.3)(7.8,0.6)
\pscoil[linewidth=0.04,coilwidth=0.2,coilheight=1.,coilarm=0.05,linecolor=red]{}(7.8,0.6)(7.8,0.)
\pscircle[linewidth=0.04,fillcolor=gray,fillstyle=solid](7.1,-0.5){0.2}
\psline[linewidth=0.02]{->}(11,2.3)(11,0.5) \rput{-90}(11.3,1.7){time}
\end{pspicture}
\caption{Schema of hadron creation via uncorrelated (left) and
  correlated (right) string breakups. Green band indicates the color
  flow ordering of the gluon ladder. Excited gluon -- which splits
  promptly into a $Q\bar{Q}$
  pair -- is marked in red.
\label{fig:induced}} 
\end{figure}
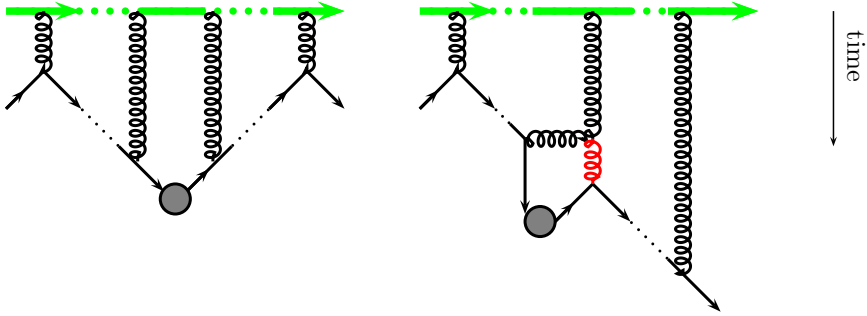

   The causal description of hadron formation turns out to be an
   excellent guiding principle in the model-building.  If we let the
   information about string breakup run along the string together
   with newly created quark,  the requirement of causal connection between
   breakup vertices implies the subsequent breakup is triggered
   by the propagating quark. Such a production mechanism seems
   suitable for narrow resonant states with mass given by
   Eq.\ref{eq:mass}. The difference between spontaneous (uncorrelated)
   and induced (correlated) string breakup is shown on diagrams in Fig.~\ref{fig:induced}.

      It is remarkable that it is the causal reformulation of the
   string fragmentation rules that actually reveals the quantized nature
   of the process. Comparing mass spectra of light
   mesons with string parametrization, it becomes clear
   that fragmentation proceeds in distant steps in helix phase
   difference,  $\Phi_B -\Phi_A=n \Delta\Phi, n=1,2,3,..$.  The
   (transverse sector of)
   quantized string can  therefore be described with help of just two
   parameters, energy scale $\kappa R$ and the phase quantum $\Delta\Phi$.
   
     The model is particularly interesting from the phenomenological
   point of view since it is empirically overconstrained: the number
   of sensitive observables vastly outnumbers the number of adjustable
   parameters. On one side, we have hadron mass spectra:   
   pseudoscalar mesons ($\pi,\eta,\eta'$) can be associated with
   quantum states ($n$=1,3,5),  vector meson $\omega$ fills the slot
   $n$=4. On the other side, the underlying helical field defines
   intrinsic transverse momenta of hadrons and, for specific
   configurations - homogeneous or slowly evolving helical field -  dominates the correlations
   between hadrons as function of their relative position along the string. 
 
     Before moving on to the discussion of phenomena which can be
   explained by the model,  a disclaimer should be issued about what
   the model does not describe - not because it is
   fundamentally defficient, but because a priority is given to a broad
   exploration of predictions stemming from the simple core concept
   of quantized fragmentation. Detailed description of $qg\rightarrow qq\bar{q}$
   interaction is not included which means
   the model is blind to the mass difference between charged and
   neutral pions - this is essentially setting the limit of the
   precision of predictions,  at the level of $\sim$3\%.  Such
   a precision is comparable with the precision of experimental
   measurements of correlations between adjacent hadrons, and
   therefore sufficient for the purpose of the current study.

\begin{figure}
\begin{center}
\includegraphics[width=0.7\textwidth]{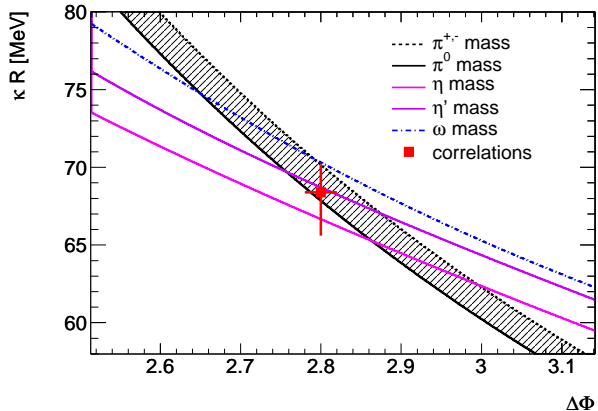}
\caption{An overview of constraints imposed on model parameters by
hadron mass spectrum~\cite{pdg} and by the measurement of correlations
between adjacent hadrons~\cite{ATLAS-chains}. The dashed region
indicates the internal precision of model predictions. 
\label{fig:1} }    
\end{center}
\end{figure}
  
     Figure~\ref{fig:1} shows the overview of constraints imposed on
     helix parameters by hadron masses~\cite{pdg} and by the
     measurement of correlations between adjacent
     hadrons \cite{ATLAS-chains} (see  Appendix \ref{appendix} for the
     extraction of parameters from the measurement). Please note
     the measured correlations cover, among other, the entire
     anomalous production of close like-sign pion pairs, traditionally
     attributed to the Bose-Einstein interference.

\begin{figure}[t]
\begin{pspicture}(0,-2.5)(10,2.5)
\pscurve[linewidth=0.1,linecolor=green]{-}(0.,2.3)(0.2,2.1)(0.5,1.9)(5.,1.7)(9.5,1.9)(10,2.)(9.5,2.1)
\pscurve[linewidth=0.1,linecolor=green,linestyle=dotted]{-}(9.5,2.1)(5.,2.2)(2.,2.1)(0.2,1.)(0.,0.)(0.2,-1.)
\psline[linewidth=0.1,linecolor=green]{-}(4.,2.)(4.5,1.7)
\psline[linewidth=0.1,linecolor=green]{-}(4.,1.4)(4.5,1.7)
\pscurve[linewidth=0.1,linecolor=green]{-}(0.2,-1)(0.5,-1.8)(5,-2.3)(9.5,-2.3)(10,-2.2)(9.5,-2.1)
\psline[linewidth=0.1,linecolor=green]{-}(4.7,-2.)(5.2,-2.3)
\psline[linewidth=0.1,linecolor=green]{-}(4.7,-2.6)(5.2,-2.3)
\pscoil[linewidth=0.04,coilwidth=0.2,coilheight=1,coilarm=0.05]{}(1.7,1.6)(1.7,0.8)
\psline[linewidth=0.04]{->}(1.2,0.8)(1.45,0.8) \psline[linewidth=0.04]{-}(1.2,0.8)(1.7,0.8)
\psline[linewidth=0.04]{->}(1.7,.8)(2.3,.8)
\pscoil[linewidth=0.04,coilwidth=0.2,coilheight=1,coilarm=0.05]{}(2.7,1.6)(2.7,0.8)
\psline[linewidth=0.04]{-}(2.2,0.8)(2.7,0.8)
\psline[linewidth=0.04]{->}(2.7,.8)(3.2,.8)
\pscoil[linewidth=0.04,coilwidth=0.2,coilheight=1,coilarm=0.05]{}(3.7,1.6)(3.7,0.8)
\psline[linewidth=0.04]{-}(3.2,0.8)(3.7,0.8)
\psline[linewidth=0.04]{->}(3.7,.8)(7.,-1.6)

\pscoil[linewidth=0.04,coilwidth=0.2,coilheight=1,coilarm=0.05]{}(5,1.6)(5,0.8)
\psline[linewidth=0.04]{->}(5.,.8)(5.5,0.8)
\pscoil[linewidth=0.04,coilwidth=0.2,coilheight=1,coilarm=0.05]{}(6,1.6)(6,0.8)
\psline[linewidth=0.04]{-}(5.4,0.8)(6,0.8)
\psline[linewidth=0.04]{->}(6.,.8)(6.5,0.8)
\pscoil[linewidth=0.04,coilwidth=0.2,coilheight=1,coilarm=0.05]{}(7,1.6)(7,0.8)
\psline[linewidth=0.04]{-}(6.4,0.8)(7,0.8)
\psline[linewidth=0.04]{->}(7.,.8)(7.5,0.8)
\pscoil[linewidth=0.04,coilwidth=0.2,coilheight=1,coilarm=0.05]{}(8,1.6)(8,0.8)
\psline[linewidth=0.04]{-}(7.4,0.8)(8,0.8)
\psline[linewidth=0.04]{->}(8.,.8)(8.5,0.8)
\pscoil[linewidth=0.04,coilwidth=0.2,coilheight=1,coilarm=0.05]{}(9,1.6)(9,0.8)
\psline[linewidth=0.04]{-}(8.4,0.8)(9,0.8)
\psline[linewidth=0.04]{->}(9.,.8)(9.5,0.8)

\pscoil[linewidth=0.04,coilwidth=0.2,coilheight=1,coilarm=0.05]{}(6.,-0.8)(6.,-1.6)
\pscoil[linewidth=0.04,coilwidth=0.2,coilheight=1,coilarm=0.05]{}(6.,-1.6)(6.,-2.4)
\pscoil[linewidth=0.04,coilwidth=0.2,coilheight=1,coilarm=0.05,linecolor=red]{}(5.,-1.6)(6.,-1.6)

\pscoil[linewidth=0.04,coilwidth=0.2,coilheight=1,coilarm=0.05]{}(4,-1.6)(4,-2.4)
\psline[linewidth=0.04]{->}(5.,-1.6)(5.,0.8)
\psline[linewidth=0.04]{-}(4.,-1.6)(5.,-1.6)
\psline[linewidth=0.04]{-}(3.4,-1.6)(4.,-1.6)
\pscoil[linewidth=0.04,coilwidth=0.2,coilheight=1,coilarm=0.05]{}(3,-1.6)(3,-2.4)
\psline[linewidth=0.04]{->}(3,-1.6)(3.5,-1.6)
\psline[linewidth=0.04]{-}(2.4,-1.6)(3.,-1.6)
\pscoil[linewidth=0.04,coilwidth=0.2,coilheight=1,coilarm=0.05]{}(2,-1.6)(2,-2.4)
\psline[linewidth=0.04]{->}(2,-1.6)(2.5,-1.6)
\psline[linewidth=0.04]{-}(1.4,-1.6)(2.,-1.6)

\pscoil[linewidth=0.04,coilwidth=0.2,coilheight=1,coilarm=0.05]{}(7,-1.6)(7,-2.4)
\psline[linewidth=0.04]{->}(7,-1.6)(7.5,-1.6)
\pscoil[linewidth=0.04,coilwidth=0.2,coilheight=1,coilarm=0.05]{}(8,-1.6)(8,-2.4)
\psline[linewidth=0.04]{->}(8,-1.6)(8.5,-1.6)
\psline[linewidth=0.04]{-}(7.4,-1.6)(8.,-1.6)
\pscoil[linewidth=0.04,coilwidth=0.2,coilheight=1,coilarm=0.05]{}(9,-1.6)(9,-2.4)
\psline[linewidth=0.04]{->}(9,-1.6)(9.5,-1.6)
\psline[linewidth=0.04]{-}(8.4,-1.6)(9.,-1.6)

\end{pspicture}
\caption{ Schema of correlated(induced) string breakup across string loops.
  Green band indicates the color
  flow ordering of the gluon ladder. Excited gluon -- which splits
  promptly into a $Q\bar{Q}$  pair  -- is marked in red.
\label{fig:loop}} 
\end{figure}
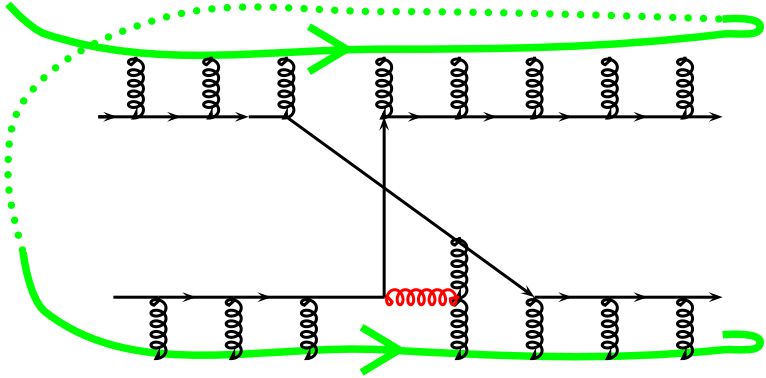

\begin{figure}[t]
\begin{pspicture}(0,-0.5)(10,2.5)
\psline[linewidth=0.05]{|->}(1,1.5)(1.5,1.5)
\psline[linewidth=0.05]{|->}(2,1.5)(3,1.5)
\psline[linewidth=0.05]{|->}(4,1.5)(5.,1.5)  \rput(7,1.5){p/n}
\psline[linewidth=0.05]{<-|}(1.5,1.)(2,1.)
\psline[linewidth=0.05]{<-|}(3,1.)(4,1.)
\psline[linewidth=0.05]{<-}(5.,1.)(6,1.) \rput(7,1.){??}
\psline[linestyle=dotted,linecolor=red](1,.5)(1,1.8)  \rput(1.2,2){\red Break}
\psline[linestyle=dotted,linecolor=red](2,.5)(2,1.8)  \rput(2.2,2){\red Break}
\psline[linestyle=dotted,linecolor=red](4,.5)(4,1.8) \rput(4.2,2){\red Break}
\psline[linestyle=dotted,linecolor=red](6,.5)(6,1.8) \rput(6.2,2){\red Break}
\psline{<->}(0,0.5)(7,0.5) 
\psline(1,0.4)(1,0.6)  \rput(1,0.2){0}
\psline(1.5,0.4)(1.5,0.6)  \rput(1.5,0.2){1}
\psline(2,0.4)(2,0.6)  \rput(2,0.2){2}
\psline(2.5,0.4)(2.5,0.6)  \rput(2.5,0.2){3}
\psline(3,0.4)(3,0.6)  \rput(3,0.2){4}
\psline(3.5,0.4)(3.5,0.6)  \rput(3.5,0.2){5}
\psline(4,0.4)(4,0.6)  \rput(4,0.2){6}
\psline(4.5,0.4)(4.5,0.6)  \rput(4.5,0.2){7}
\psline(5,0.4)(5,0.6)  \rput(5,0.2){8}
\psline(5.5,0.4)(5.5,0.6)  \rput(5.5,0.2){9}
\psline(6,0.4)(6,0.6)  \rput(6,0.2){10}
           \rput(7,0.2){[$\Delta\Phi$]}
\end{pspicture} 
\begin{center}
\includegraphics[width=0.7\textwidth]{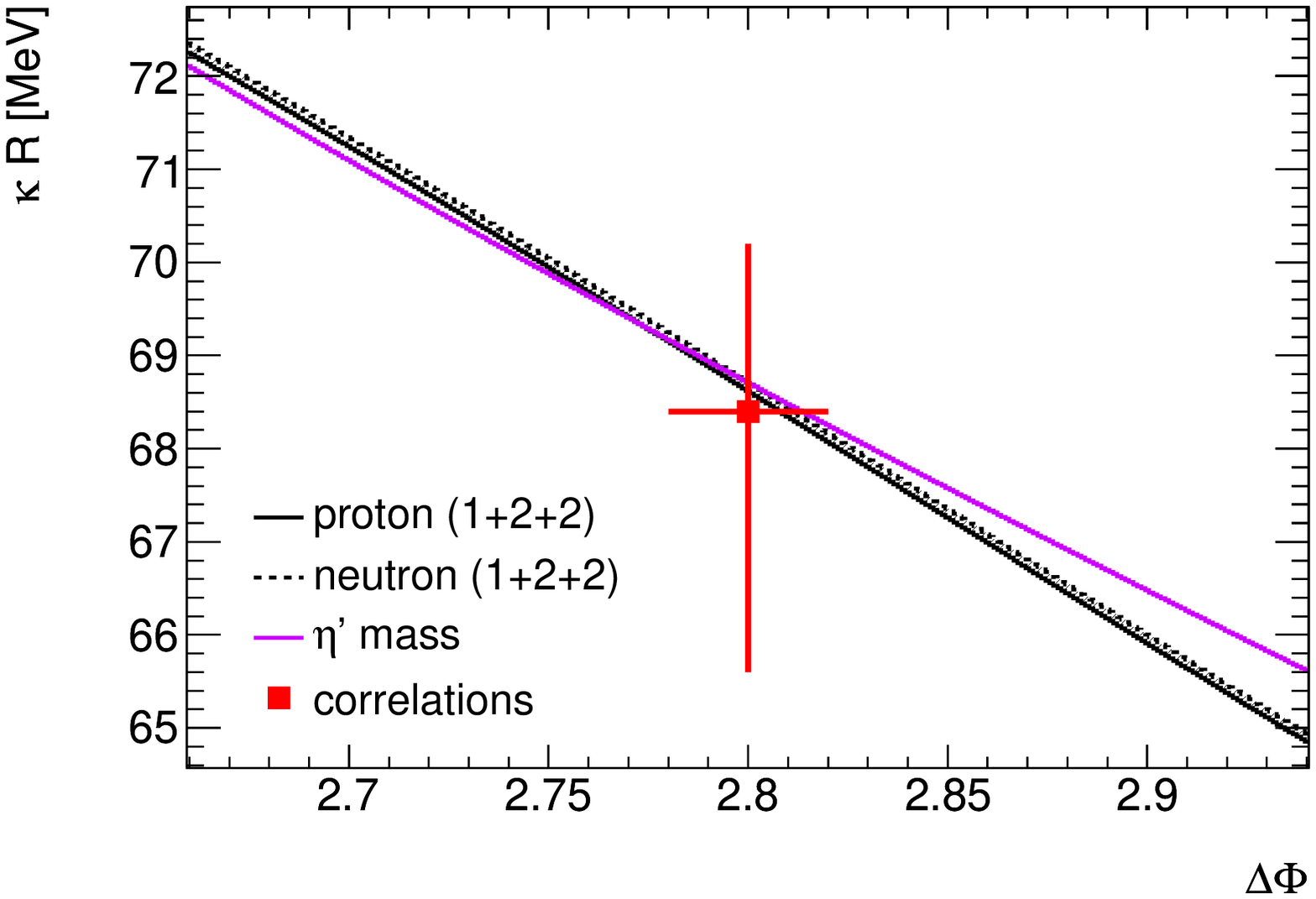}
\caption{Top: Example of string breakup leading to creation of a nucleon
  (2+2+1) in forward direction.  Bottom: Constraints imposed on model parameters by (2+2+1) nucleon
  hypothesis, compared with those stemming from $\eta'$ mass and 
  from the measured correlations between adjacent hadrons~\cite{ATLAS-chains}. 
\label{fig:m122} }  
\end{center}
\end{figure}

 \section{Baryon production}  \label{sec:baryons}

     In the introduction, the causal connection between string breakup
     vertices was limited to the information transfer along the
     string. It is however possible that such an transfer 
     occurs across the string loops, in particular for string
     configuration with dense packing (small pitch). This type of induced
     string breakup  is shown in Fig.~\ref{fig:loop}. The
     net result of such an interaction is a coherent production of three
     partons ($qQ\bar{Q}$ or $\bar{q}Q\bar{Q}$, where $q$ and
     $\bar{q}$ indicate the propagating parton which triggers the
     interaction). For the production of a pair of baryons, it is therefore
     sufficient to consider an initial string breakup ($g\rightarrow
     q\bar{q}$) followed by two induced gluon splittings across string
     loops.  The integration over longitudinal component of the string field
     contributes to the separation of quark and antiquark momenta
     ( running forward and backward along the string), and this may
     facilitate creation of bound states composed of colour neutral
     combinations of quarks or antiquarks only, the baryons.    

       The lightest baryons (nucleons) have masses which slightly exceed
     mass of  two string loops ( $4\pi\kappa R\sim$ 0.9 GeV, using
     the measurement~\cite{ATLAS-chains} ). The proximity with mass of
     $\eta'$ suggests that these particles are formed by 5 quanta of
     quantized helical string. Several possible production
     mechanisms are investigated below. They differ by the redistribution of
    initial momentum between partons.  Please note that impact
    of quark masses on the mass of hadrons diminishes as hadrons
    become heavier : the mass difference between neutron and proton
    is only 1.3\textperthousand.  If the model is capturing the
    dynamics of hadron formation correctly, we may also expect the precision
    of model predictions to improve for heavier hadrons.

\subsection{Case (1+2+2)}

  The first example of possible string breaking scheme
for a nucleon formation is
$q(\bar{q})q(\bar{q})q(\bar{q})$=1(1)2(2)2(2) 
 - this compact notation is used to describe the forward/$qqq$ and
(backward/$\bar{q}\bar{q}\bar{q}$) baryon simultaneously.
 The break-up pattern and the mass constraints on model parameters corresponding to nucleon moving
 in forward direction ($qqq$, by convention) are shown in
 Fig.~\ref{fig:m122}(string breakup is modelled via gluon
 splitting $g\rightarrow q\bar{q}$).  The nucleon mass dependence on
model parameters nearly coincides with the curve calculated for
$\eta'$ meson, which is very encouraging : the model is working
equally well for baryons and mesons. It seems, nevertheless, that this
solution is not quite optimal, because the forward and
backward configurations are not symmetric and
their masses differ by $\sim$2\%. 
   
\begin{figure}[h]
\begin{pspicture}(0,-0.5)(10,4)
\psline[linewidth=0.05]{|->}(1,3.5)(2.,3.5)
\psline[linewidth=0.05]{|->}(3.,3.5)(3.5,3.5)
\psline[linewidth=0.05]{|->}(4,3.5)(5.,3.5)  \rput(7,3.5){p/n}
\psline[linewidth=0.05]{<-|}(2.,3.)(3,3.)
\psline[linewidth=0.05]{<-|}(3.5,3.)(4,3.)
\psline[linewidth=0.05]{<-|}(5.,3.)(6,3.) \rput(7,3.){$\bar{p}/\bar{n}$}
\psline[linestyle=dotted,linecolor=red](1,3.)(1,3.8)  \rput(1.2,4.){\red Break}
\psline[linestyle=dotted,linecolor=red](3,3.)(3,3.8)  \rput(3.2,4.){\red Break}
\psline[linestyle=dotted,linecolor=red](4,3.)(4,3.8) \rput(4.2,4.){\red Break}
\psline[linestyle=dotted,linecolor=red](6,3.)(6,3.8) \rput(6.2,4.){\red Break}
\rput{90}(0.2,3.5){(2+1+2)}
\psline[linewidth=0.05]{|->}(1,1.5)(1.5,1.5)
\psline[linewidth=0.05]{|->}(2,1.5)(3.5,1.5)
\psline[linewidth=0.05]{|->}(5,1.5)(5.5,1.5)  \rput(7,1.5){p/n}
\psline[linewidth=0.05]{<-|}(1.5,1.)(2,1.)
\psline[linewidth=0.05]{<-|}(3.5,1.)(5,1.)
\psline[linewidth=0.05]{<-|}(5.5,1.)(6,1.) \rput(7,1.){$\bar{p}/\bar{n}$}
\psline[linestyle=dotted,linecolor=red](1,.5)(1,1.8)  \rput(1.2,2){\red Break}
\psline[linestyle=dotted,linecolor=red](2,.5)(2,1.8)  \rput(2.2,2){\red Break}
\psline[linestyle=dotted,linecolor=red](5,.5)(5,1.8) \rput(5.2,2){\red Break}
\psline[linestyle=dotted,linecolor=red](6,.5)(6,1.8) \rput(6.2,2){\red Break}
\rput{90}(0.2,1.5){(1+3+1)}
\psline{<->}(0,0.5)(7,0.5) 
\psline(1,0.4)(1,0.6)  \rput(1,0.2){0}
\psline(1.5,0.4)(1.5,0.6)  \rput(1.5,0.2){1}
\psline(2,0.4)(2,0.6)  \rput(2,0.2){2}
\psline(2.5,0.4)(2.5,0.6)  \rput(2.5,0.2){3}
\psline(3,0.4)(3,0.6)  \rput(3,0.2){4}
\psline(3.5,0.4)(3.5,0.6)  \rput(3.5,0.2){5}
\psline(4,0.4)(4,0.6)  \rput(4,0.2){6}
\psline(4.5,0.4)(4.5,0.6)  \rput(4.5,0.2){7}
\psline(5,0.4)(5,0.6)  \rput(5,0.2){8}
\psline(5.5,0.4)(5.5,0.6)  \rput(5.5,0.2){9}
\psline(6,0.4)(6,0.6)  \rput(6,0.2){10}
           \rput(7,0.2){[$\Delta\Phi$]}
\end{pspicture} 
\begin{center}
\includegraphics[width=0.7\textwidth]{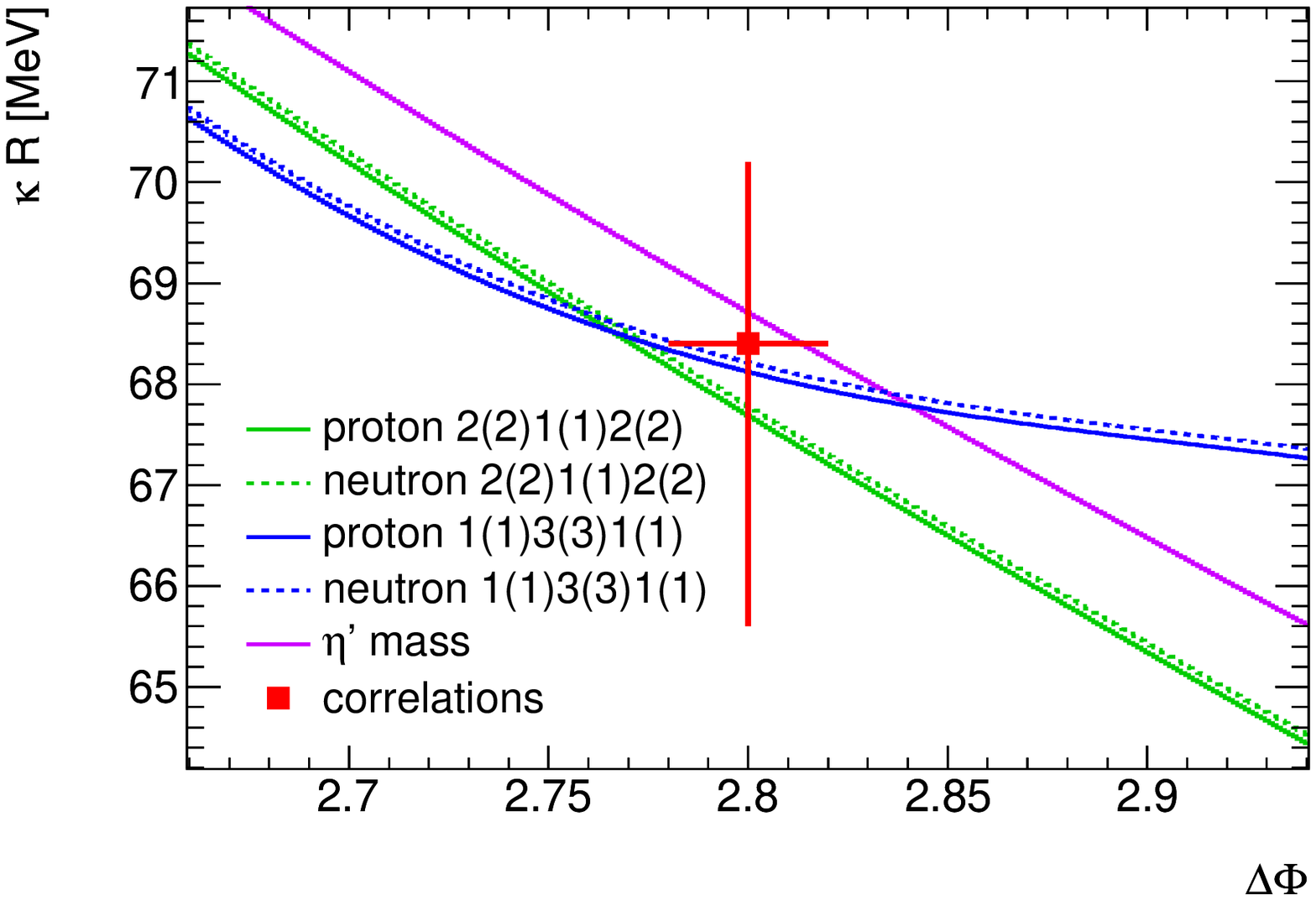}
\caption{Top: (2+1+2)  and (1+3+1)  nucleon formation scenarios,
  symmetrized with respect to ($qqq$/$\bar{q}\bar{q}\bar{q}$)
  production. Bottom: Constraints imposed on model parameters by 
  these nucleon production scenarios, compared with those stemming from $\eta'$ mass and 
  from the measured correlations between adjacent hadrons~\cite{ATLAS-chains}. 
\label{fig:symm} }  
\end{center}
\end{figure}

\subsection{Symmetrization (2+1+2) \& (1+3+1) }

    The string breaking pattern can be symmetrized with respect to $qqq$
    / $\bar{q}\bar{q}\bar{q}$ production as shown in
    Fig.~\ref{fig:symm}  for (2+1+2) and (1+3+1) nucleon production
    scenarios. In both cases, we obtain a 
    baryon/antibaryon pair  with parametrization
    matching the meson production and correlation measurements, 
    within $\sim$1\% precision tolerance.

\subsection{$\Lambda$-like configurations }

\begin{figure}[t]
\begin{center}
\begin{pspicture}(0,-0.5)(12,4)
\psline[linewidth=0.05]{|->}(1,3.5)(2.,3.5)
\psline[linewidth=0.05]{|->}(2.5,3.5)(4.,3.5)
\psline[linewidth=0.05]{|->}(5.5,3.5)(6.,3.5)  \rput(7.5,3.5){$\Lambda$-like}
\psline[linewidth=0.05]{<-|}(2.,3.)(2.5,3.)
\psline[linewidth=0.05]{<-|}(4.,3.)(5.5,3.)
\psline[linewidth=0.05]{<-|}(6.,3.)(7.,3.) \rput(7.5,3.){$\bar{\Lambda}$-like}
\psline[linestyle=dotted,linecolor=red](1,2.5)(1,3.8)  \rput(1.2,4){\red Break}
\psline[linestyle=dotted,linecolor=red](2.5,2.5)(2.5,3.8)  \rput(2.7,4){\red Break}
\psline[linestyle=dotted,linecolor=red](5.5,2.5)(5.5,3.8) \rput(5.7,4){\red Break}
\psline[linestyle=dotted,linecolor=red](7,2.5)(7,3.8) \rput(7.2,4){\red Break}
\rput{90}(0.2,3.5){(1+3+2)}
\psline[linewidth=0.05]{|->}(1.5,1.5)(2.,1.5)
\psline[linewidth=0.05]{|->}(2.5,1.5)(4.,1.5)
\psline[linewidth=0.05]{|->}(5.5,1.5)(6.,1.5)  \rput(7.5,1.5){$\Lambda$-like}
\psline[linewidth=0.05]{|->}(6.5,1.5)(7.,1.5)
\psline[linewidth=0.05]{<-|}(1.,1.)(1.5,1.)
\psline[linewidth=0.05]{<-|}(2.,1.)(2.5,1.)
\psline[linewidth=0.05]{<-|}(4.,1.)(5.5,1.)
\psline[linewidth=0.05]{<-|}(6.,1.)(6.5,1.) \rput(7.5,1.){$\bar{\Lambda}$-like}
\psline[linestyle=dotted,linecolor=red](1,.5)(1,1.8)  \rput(0.8,2){\red Break}
\psline[linestyle=dotted,linecolor=red](1.5,.5)(1.5,1.8)  \rput(1.7,1.8){\red Break}
\psline[linestyle=dotted,linecolor=red](2.5,.5)(2.5,1.8)  \rput(2.7,2){\red Break}
\psline[linestyle=dotted,linecolor=red](5.5,.5)(5.5,1.8) \rput(5.5,2){\red Break}
\psline[linestyle=dotted,linecolor=red](6.5,.5)(6.5,1.8) \rput(6.5,1.8){\red Break}
\psline[linestyle=dotted,linecolor=red](7,.5)(7,1.8) \rput(7.2,2){\red Break}
\rput{90}(0.2,1.4){(1+3+1+1)}
\psline{<->}(0,0.5)(8,0.5) 
\psline(1,0.4)(1,0.6)  \rput(1,0.2){0}
\psline(1.5,0.4)(1.5,0.6)  \rput(1.5,0.2){1}
\psline(2,0.4)(2,0.6)  \rput(2,0.2){2}
\psline(2.5,0.4)(2.5,0.6)  \rput(2.5,0.2){3}
\psline(3,0.4)(3,0.6)  \rput(3,0.2){4}
\psline(3.5,0.4)(3.5,0.6)  \rput(3.5,0.2){5}
\psline(4,0.4)(4,0.6)  \rput(4,0.2){6}
\psline(4.5,0.4)(4.5,0.6)  \rput(4.5,0.2){7}
\psline(5,0.4)(5,0.6)  \rput(5,0.2){8}
\psline(5.5,0.4)(5.5,0.6)  \rput(5.5,0.2){9}
\psline(6,0.4)(6,0.6)  \rput(6,0.2){10}
\psline(6.5,0.4)(6.5,0.6)  \rput(6.5,0.2){11}
\psline(7,0.4)(7,0.6)  \rput(7,0.2){12}
           \rput(8,0.2){[$\Delta\Phi$]}
\end{pspicture} 
\includegraphics[width=0.7\textwidth]{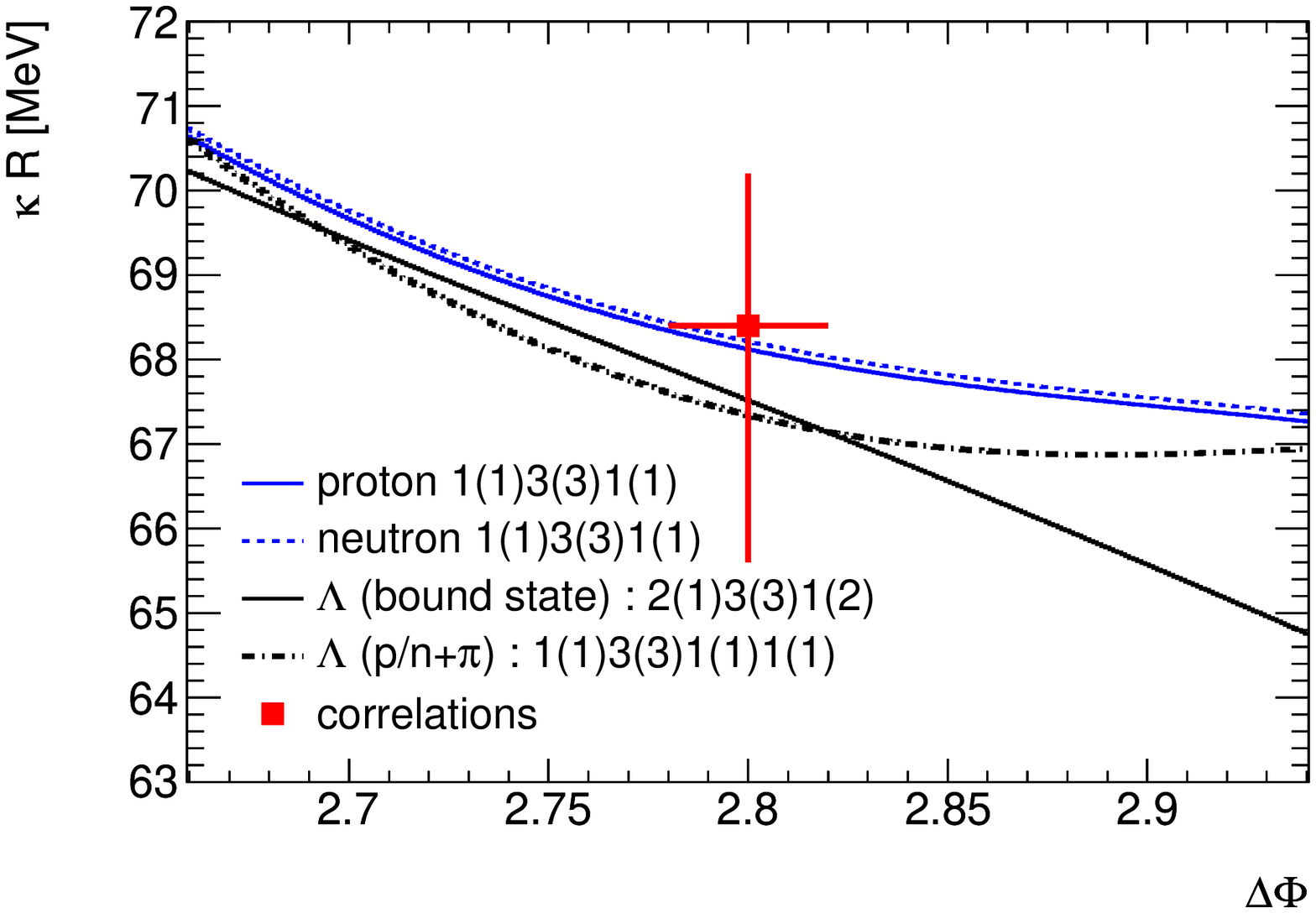}
\caption{Top: $\Lambda\bar{\Lambda}$-like string splitting scenarios,
  bound (1+3+2) and unbound (1+3+1+1)=(p+$\pi$).  
  Bottom: Constraints imposed on model parameters by requiring that these 
  configurations acquire mass of $\Lambda$ baryon, compared with those stemming
  from nucleon mass and 
  from the measured correlations between adjacent hadrons~\cite{ATLAS-chains}. 
\label{fig:lambda} }
\end{center}  
\end{figure}

  With a simple modification of the nucleon breaking pattern, a
  $\Lambda$-like baryon can be produced. An example is shown
  in Fig.~\ref{fig:lambda} using 2(1)3(3)1(2) split pattern which produces
  a pair of baryons with $\Lambda$ mass. Even more intriguingly,
  $\Lambda$ production can be mimicked by a simple chain production
  of proton and pion, where the unbound state also ends up with mass
  of $\Lambda$.  Without a detailed production mechanism
  for $s\bar{s}$ introduced in the model (subject of a separate
  publication), the discussion of $\Lambda$ production is
  premature, however it seems that (a) it is very likely that
  strangeness can be introduced in the model without a major
  perturbation of the string parametrization, (b) $\Lambda$ production
  can be to some extent mimicked by baryon production involving
  light(massless) quark pairs only, or even by unbound production
  of ($p/n+\pi$) pair with rank difference 2 (the rank of direct
  hadrons refers to their ordering along the fragmenting string,
  according to the colour flow).

\section{Discussion: mass spectra of light hadrons}
\begin{figure}[t]
\begin{center}
\includegraphics[width=0.8\textwidth]{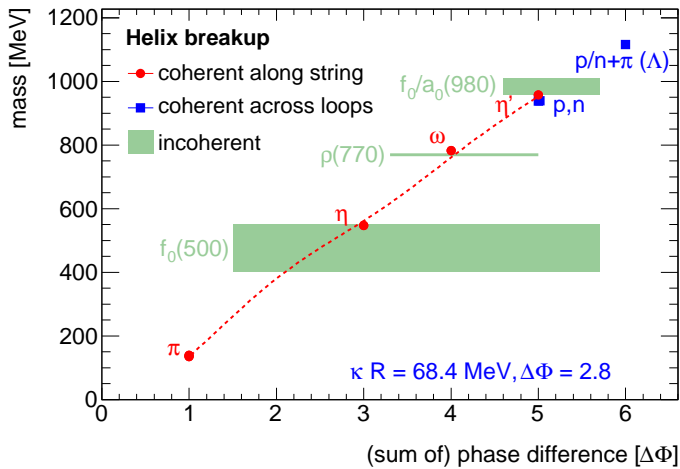}
\caption{Light hadrons classified according to presumed production
  mechanism in the model of helical QCD string, plotted as function of the (quantized) helix phase
  difference between breakup vertices. For baryons, the phase
  difference is summed over three string pieces. Coherent string
  breakup defines the hadron mass (red dashed line) as function of
  transverse string parameters, wide resonant states
  (f$_0$,$\rho$)  have mass influenced by the longitudinal string
  properties. Calculations done with string parameters ($\kappa R,\Delta\Phi$) measured 
  via hadron correlations~\cite{ATLAS-chains}.
\label{fig:spectrum} }  
\end{center}
\end{figure}

   With the extension of the model to the baryon production - which turns
   out to be simply a consistency check, since nucleons appear
   just where one would expect them, when  
   interactions across string loops are admitted - the matching of
   light hadrons to the quantized fragmentation scheme is nearly complete.
   In Fig.~\ref{fig:spectrum},  mass spectrum is plotted
   as function of the number of string quanta involved in hadron
   formation. Colors are used to distinguish set of hadrons with
   similar production mechanism: in red, pseudoscalar mesons
   ($\pi,\eta,\eta'$) and vector meson $\omega$ for coherent
   string ``light front'' breaking. The red dashed line indicates
   the (non-quantized) mass dependence on the helix phase difference.
   Baryons produced via coherent string breaking across string loops 
   are shown in blue.  Wide resonant states ($f_0,\rho$), in green, are presumably
   produced via incoherent string breaking: in such a scenario, mass
   of hadron is not uniquely defined by the tranverse shape of the
   string, it also depends on longitudinal properties of the string. 
   Indeed, the $\rho$ mass measurements differ in $\tau$ decays
   and hadroproduction \cite{pdg}.  

   Having the light hadron mass spectrum and part of hadron correlations described by just two
   parameters with a precision of few percent is certainly an
   attractive feature of the model, allowing to reduce the
   number of parameters needed for description of data.
   It is also an opportunity to reconsider the approach to soft QCD,
   mostly because it seems
   difficult to reconcile the quantized fragmentation with the
   conventional picture of colour field consisting of
   ( nonperturbative ) sea of soft gluons. What would prevent any of
   these gluons to produce a random breakup, which would then prevail
   over coherent string breaking ?  In fact, the arguments tend to
   support just the opposite - a minimalist description of string
   with density of gluons as low
   as two per $\Delta\Phi$. For the argumentation we do not need to
   reach further than the original helix paper~\cite{helix}. One of the main arguments of the authors was the
   effective absence of collinear gluon emission at the end of the
   parton cascade. They evaluated the minimal distance between colour connected
   gluons to be 11/6 (1.83),  which is just 8\% smaller than the measured
   value $2\sin{(\Delta\Phi/2)}=$1.97. The space between gluons,
   according to the discussion in \cite{helix}, is effectively
   occupied by the accompanying Coulomb field. For a creation of
   lightest hadron ($qg\bar{q}$) , at least one additional
   accompanying gluon is needed in each $\Delta\Phi$ interval.  An
   effective density of 1 gluon per $\Delta\Phi/2\sim$1.4 is also roughly
   consistent with the non-collinear emission of gluons by quarks, where the 
   effective depleted zone around quark is 3/2.  In the original
   paper, authors did not deploy the causal constraint, and concentrated
   on the parametrization of the longitudinal separation of partons in
   $\Delta y$ instead. The quantized fragmentation is extending the
   argument to the limit of $\Delta y\rightarrow$0 by fully exploiting
   the azimuthal degree of freedom.

\section{Further experimental evidence}
    In the model of quantized fragmentation of a helical QCD string,
  transverse momenta of hadrons (w.r.t. string axis) are
  constrained and predicted with a similar precision as their masses.
  Since the measurement of these intrinsic transverse momenta is
  conditioned by the precision with which the axis of the generating
  QCD string can be determined, the measurement is not expected
  to improve the precision of constraints derived from hadron mass
  spectrum and from measurement of correlations between hadrons
  as described above. Nevertheless, study of pencil-like events
  in LEP data or soft minimum bias events at LHC may provide
  an interesting cross-check of model predictions, which are gathered
  in Table~\ref{tab:pt}. All calculations are done assuming ideal
  helix shape without additional complex structures (knots).

\begin{table}[h!]{
\caption{Model prediction for intrinsic transverse momentum of direct
  hadrons, for measured (Appendix~\ref{appendix}) string parameters. \label{tab:pt}}
\begin{center}
\begin{tabular}{|l|c|c|c|}
\hline
hadron &  production mode & quantized content [$\Delta\Phi$] & $p_T$ [MeV] \\
\hline
$\pi$       &  induced, light-front &  1  &   135 (+4,-6) \\
$\eta$      & induced, light-front &  3  &   119 (+3,-5) \\
$\omega$  &  induced, light-front &  4 &  86 (+2,-4)  \\
$\eta'$      & induced, light-front & 5 &  90 (+3,-4) \\
p,n     &  induced, across loop &  1+2+2 &  206 (+6,-9) \\
p,n     &  induced, across loop &  2+1+2 &  135 (+4,-6) \\
p,n     &  induced, across loop &  1+3+1 &  172 (+5,-7) \\
$\Lambda$ & induced, across loop & 1+3+2 &  208 (+6,-9) \\    
\hline
\end{tabular}
\end{center}}
\end{table}

\section{Conclusions}
    The production of light baryons using the hypothesis of correlated
    helical string breaking with information passing across helix
    loops has been investigated.  Nucleons (p,n) can be described by a
    combination of three coherently produced fragments of helical
    string, which jointly carry 5 string quanta. Model is using ideal helix shape and massless parton
    approximation. Despite the model simplicity, masses of light
    hadrons below 1 GeV agree with model predictions within 3\% 
    and the agreement is even better ($\sim$1\%) for baryons.
    No addditional parameters nor readjustment of string
    parameterization are needed to accommodate the baryon production.

      It is argued that the ideal helix approximation can be a proxy
    for a low density gluon chain, with approximately
    two gluons per $\Delta\Phi$. Such a simplified description of the
    confining field should be of interest for hadronization models 
    and for calculations of parton density functions (PDF) within
    nucleon.

\section*{Acknowledgements}
This work is supported by the Inter-Excellence/Inter-Transfer grant LT17018 and the Research Infrastructure project LM2018104
 funded by Ministry of Education, Youth and Sports of the Czech Republic, and the Charles University project UNCE/SCI/013.

\appendix
\section{Model parameters measured with help of hadron correlations} \label{appendix}

   Correlations between pairs of hadrons(pions) with rank difference 1
   and 2 have been
   measured~\cite{ATLAS-chains} in hadron chains selected
   via mass minimization and matched with the
   anomalous production of like-sign (LS) pairs. Charge ordering of hadrons
   implies opposite-sign (OS) pairs correspond to rank difference 1,
   and LS pairs to rank difference 2. The measured
   value of momentum difference is (266$^{+8}_{-11}$) MeV for OS pairs,  and
 (89.7$^{+2.5}_{-3.5}$) MeV for LS pairs.

     In the limit of homogeneous string field (constant helix pitch),
    the  momentum difference between pion pairs with rank difference $n$ can
     be written as~\cite{helix2}  
    \begin{equation}
     Q(n)  =  2 \ p_\mathrm{T}(n=1) \ \sin{ (n \Delta\Phi / 2)}   \\       
\end{equation}

  The ratio
\begin{equation} 
 \frac{Q(n=1)}{Q(n=2)} =\frac{\sin{(\Delta\Phi / 2)}}{\sin{(\Delta\Phi)}}
   = 2.96 \pm 0.15
\end{equation}
   implies $ \Delta\Phi  = 2.80 \pm 0.02$.  Taking into account
  Eq.\ref{eq:calc}, the energy scale estimate is $\kappa R  =  68.4 ^{+1.8}_{-2.8}$ GeV .

 \end{document}